\documentstyle[preprint,aps,prl,epsf]{revtex}
\begin{document}
\title{Dynamic structure factors of a dense mixture}
\author{Supurna Sinha}
\address{Jawaharlal Nehru Centre for Advanced Scientific Research, 
Bangalore, 560012 India\\
{\rm and} Department of Physics, Indian Institute of 
Science, Bangalore, 560012 India}
\maketitle
\widetext

\begin{abstract}
We compute the dynamic structure factors of a dense binary liquid mixture. 
These describe dynamics on molecular length scales, where structural 
relaxation is important. We find that the presence of a few large particles 
in a dense fluid of small particles slows down the dynamics considerably. 
We also observe a deep narrowing of the spectrum for a disordered mixture 
composed of a nearly equal packing of the two species. In contrast, a few 
small particles diffuse easily in the background of a dense fluid of large 
particles. We expect our results to describe neutron scattering from a dense 
mixture. \\~\\
\end{abstract}

\section{INTRODUCTION}

\noindent
The behavior of a fluid at large length and time scales is well described by 
hydrodynamics. In recent years hydrodynamics has been extended to molecular 
length scales [1]. This generalized hydrodynamic theory (GHT) was motivated 
by the desire to describe the neutron- scattering spectrum 
$S(k,\omega)$ of a dense liquid. $S(k,\omega)$, which reflects the 
molecular scale dynamics of the fluid, has a 
central diffusive peak which dramatically narrows with increasing density 
[2]. GHT provides a physically appealing description of this phenomenon. 
Structural relaxation is very slow on molecular scales in a dense liquid 
due to close packing of the molecules. Thus density fluctuations decay 
very slowly via self diffusion at $k \sim k_{0}$ where $S(k)$ has its 
maximum. The emergence of this long-lived mode leads to a reduction 
in the half-width of the scattering spectrum. \\

\noindent
In the past few years, much attention has been focused on the study of the 
dynamical properties of dense mixtures [3]. A dense binary mixture is 
known to be a better glass former than a one component liquid. Therefore, an 
understanding of the dynamics of a dense binary mixture is of considerable 
interest in analyzing the slow relaxation process that ultimately leads to 
the glass transition. \\

\noindent
Recently, the intermediate wave-vector dynamics of a dense binary 
hard-sphere mixture has been studied [4] in terms of the density 
fluctuations of the two species. The main approximations that have gone into 
this study are the following. First of all, interdiffusion has been 
neglected. Secondly, the wave-vector dependence of the self-diffusion 
coefficients has been neglected. \\

\noindent
A more detailed analysis of a dense binary hard-sphere mixture has been 
carried out in Ref. [5]. Here, the full linear generalized hydrodynamics has 
been solved and it has been explicitly demonstrated that the density 
fluctuations are the only slow fluctuations on molecular length scales. In 
addition, both self-diffusion and interdiffusion effects with their 
wave-vector dependence, have been retained. Interdiffusion is a physical 
process special to a mixture, not shared by a one-component fluid. 
Therefore, inclusion of interdiffusion elicits the special characteristics 
of the dynamics of a mixture. The wave-vector dependence of the 
transport coefficients brings out the nonlocal effects arising from the 
short-ranged interaction potential in a dense liquid. Furthermore, the 
analysis presented in [5] incorporates certain additional static couplings 
in the extended hydrodynamic equations which had been neglected earlier.\\

\noindent
In the present paper, we use the the density modes obtained in [5] to 
evaluate the dynamic structure factors of a dense binary mixture. We expect 
the dynamic structure factors presented here to describe the molecular-scale 
dynamics of a dense liquid mixture measured by neutron scattering. \\

\noindent
The paper is organized as follows. In Sec. II we present a theoretical 
derivation of the partial dynamic structure factors of a dense mixture. In 
Sec. III we numerically calculate $S(k,\omega)$ for a few specific 
values of the parameters of the mixture and make some predictions. 
Finally, in Sec. IV we end this paper with a few concluding remarks.\\

\section{THEORETICAL DERIVATION }

\noindent
The differential scattering cross section of a binary mixture measured in a 
neutron-scattering experiment is proportional to the dynamic structure 
factor $S(k,\omega)$ given by [6] 

\begin{eqnarray}
S(k,\omega) =&& x_{1}b^{2}S_{11} (k,\omega) + x_{2}S_{22} (k,\omega)\nonumber\\
            &&+ 2\sqrt{x_{1}x_{2}}bS_{12} (k,\omega),
\end{eqnarray}
where $x_{i} = n_{i}/(n_{1} + n_{2})$ is the relative number concentration 
in the mixture of particles of component $i \cdot b$ is the ratio of the 
scattering length of the first component to that of the second. The partial 
dynamic structure factors $S_{ij} (k,\omega) (i,j = 1,2)$ appearing in 
(1) are given by 

\begin{equation}
S_{ij} (k,\omega) = \frac{1}{2\pi} \int^{\infty}_{-\infty} dt\;\;e^{i\omega t} F_{ij} (k,t).
\end{equation}
These are the temporal Fourier transforms of the time correlation functions 
$F_{ij} (k,t)$ defined as 

\begin{eqnarray*}
F_{ij} (k,t) = \langle \delta n^{*}_{i\vec{k}} (0) \delta n_{j\vec{k}} (t)\rangle.
\end{eqnarray*}

\noindent
The brackets $\langle\rangle$ indicate an equilibrium grand canonical 
ensemble average. 
$\delta n_{i\vec{k}} (i = 1,2)$, the fluctuation of the number density 
of the $i{\rm th}$ component at wave-vector $\vec{k} \neq \vec{0}$ 
is given by 

\begin{eqnarray*}
\delta n_{i\vec{k}} = \frac{1}{\sqrt{N_{i}}} \sum^{N_{i}}_{a=1} 
e^{i\vec{k} \cdot \vec{r}^{(i)}_{a}} \nonumber
\end{eqnarray*}
for a fluid containing $N_{i}$ of type $i$ particles. $a$ labels the 
particle number. \\

\noindent
It has been shown in [5] that momentum and temperature fluctuations decay 
very fast on molecular length scales in a dense binary liquid mixture. 
Therefore the molecular-scale hydrodynamic description of such a system 
involves only the partial density fluctuations of the two species, which are 
long lived on such length scales. These fluctuations are the intermediate 
wave-vector extensions of the long-wavelength heat and diffusion modes. \\

\noindent
It is convenient to consider linear combinations 
$a_{\mu\vec{k}} (\mu = 1, \Delta)$ of the number 
density fluctuations $\delta n_{1\vec{k}}$ and $\delta n_{2\vec{k}}$
that form an orthonormal set [5]. We choose  

\begin{eqnarray*}
a_{1\vec{k}} = \frac{\delta n_{1\vec{k}}}{\sqrt{n_{1}S_{11}(k)}}\nonumber
\end{eqnarray*}
and

\begin{eqnarray*}
a_{\Delta\vec{k}} =&& \frac{1}{\cos\alpha (k)} \left[\sin\alpha (k) 
\frac{\delta n_{1\vec{k}}}{\sqrt{n_{1}S_{11}(k)}}\right.\\
&&- \left.\frac{\delta  n_{2\vec{k}}}{\sqrt{n_{2}S_{22}(k)}}\right].
\end{eqnarray*}
Here $\alpha (k)$ is given by

\begin{eqnarray*}
\alpha (k) = \sin^{-1} \left[\frac{S_{12}(k)}{\sqrt{S_{11}(k)S_{22}(k)}}\right],\nonumber
\end{eqnarray*}
with $S_{ij} (k) (i,j = 1,2)$ representing the partial static
structure factors of the mixture [5]. The time evolution of the densities 
$a_{1\vec{k}}$ and $a_{\Delta\vec{k}}$ is governed by the pseudo-Liouville 
operator [5]. The Laplace transform 
of the generalized hydrodynamic equations for the number density 
fluctuations of a dense binary liquid mixture may be written in the 
following compact matrix form: 

\begin{equation}
[z{\bf\rm 1} + {\bf\rm A}(k_{1},z)]|a(\vec{k}, z)\rangle = |a(\vec{k}, t = 0)\rangle ,
\end{equation}
where $|a(\vec{k},z)\rangle$ represents a two component column vector 
with $a_{1\vec{k}}(z)$ and $a_{\Delta\vec{k}}(z)$ as its entries. 
$a_{i\vec{k}}(z) (i = 1,\Delta )$ are expressed as

\begin{eqnarray*}
a_{i\vec{k}}(z) = \int^{\infty}_{0} dt\; e^{-(z-L_{+})t}a_{i\vec{k}} = 
\frac{a_{i\vec{k}}}{(z - L_{+})}
\end{eqnarray*}
for ${\rm Re}(z) > 0$, with $L_{+}$ the pseudo-Liouville operator 
defined as in [5]. The hydrodynamic matrix ${\bf A}(k,z)$ has the 
following elements: 

\begin{eqnarray*}
A_{11} (k,z) &=& M_{11}(k,z) - \frac{\Omega_{1l}(k)^{2}}{\Omega_{ll}(k)},\\
A_{1\Delta}(k,z) &=& M_{1\Delta}(k,z) - 
\frac{\Omega_{1l}(k)\Omega_{\Delta l}(k)}{\Omega_{ll}(k)}\\
&=& A_{\Delta 1}(k,z),
\end{eqnarray*}
and
\begin{eqnarray*}
A_{\Delta\Delta}(k,z) = M_{\Delta\Delta}(k,z) - 
\frac{\Omega_{\Delta l}(k)^{2}}{\Omega_{ll}(k)}.
\end{eqnarray*}
Here $M_{ij}(k, z)$ with $i,j = 1,\Delta$ and $\Omega_{ij}(k)$ with 
$i = 1, \Delta ,l$ and $j = l$ are given in [5]. The solution to (3) can be
written as 

\begin{equation}
|a(\vec{k},z)\rangle = {\bf R}(k,z)[z{\bf 1} + {\bf A}^{d}(k,z)]^{-1}{\bf R}^{T}(k,z)|a(\vec{k}, 0)\rangle .
\end{equation}

\noindent
In Eq. (4) ${\bf A}^{d}(k,z) = {\bf R}^{T}(k,z) {\bf A}(k,z){\bf R}(k,z)$ 
is a diagonal matrix with the eigenvalues

\begin{eqnarray*}
z \pm = - \frac{[A_{11} (k,z) + A_{\Delta\Delta}(k,z)] \pm \sqrt{[A_{11}(k,z) - A_{\Delta\Delta} (k,z)]^{2} + 4A_{1\Delta} (k,z)^{2}}}{2}\nonumber
\end{eqnarray*}
of the matrix ${\bf A}(k,z)$ as its diagonal elements. The eigenfunctions 
$|\Theta_{\alpha}\rangle(\alpha = \pm)$ of the hydrodynamic matrix 
${\bf A}(k,z)$ are column vectors with 
$\frac{A_{1\Delta}(k,z)}{\sqrt{A_{1\Delta}(k,z)^{2} + [A_{11}(k,z) - z_{\alpha}]^{2}}}$ and $\frac{[z_{\alpha} - A_{11}(k,z)]}{\sqrt{A_{1\Delta}(k,z)^{2} + [A_{11}(k,z) - z_{\alpha}]^{2}}}$ as their entries. These eigenfunctions form 
the two columns of the orthogonal matrix ${\bf R}(k,z)$ which diagonalizes
${\bf A}(k,z)$.\\

\noindent
Equation (4) can therefore be recast as

\begin{equation}
|\delta n(\vec{k},z)\rangle = {\bf G}(k,z)|\delta n(\vec{k},0)\rangle ,
\end{equation}
where $|\delta n(\vec{k}, z)\rangle$ is a column vector with 
$\delta n_{i\vec{k}}(z) (i = 1,2)$ as its
entries and the matrix ${\bf G}(k,z)$ is given by

\begin{eqnarray*}
{\bf G}(k,z) = &&{\bf B}^{-1}(k,z) {\bf R}(k,z)\\
               &&\times [z{\bf 1} + {\bf A}^{d}(k,z)]^{-1}{\bf R}^{T}(k,z){\bf B}(k,z).
\end{eqnarray*}

\noindent
Here ${\bf B}(k,z)|\delta n(\vec{k},z)\rangle = |\delta a(\vec{k},z)\rangle$.
In terms of components Eq. (5) reads 

\begin{equation}
\delta n_{i} (\vec{k},z) = G_{ij} (k,z) \delta n_{j} (\vec{k},t = 0).
\end{equation}
Multiplying both sides of Eq. (6) by $\delta n_{l} (-\vec{k},t = 0)$ 
and taking a thermal average we obtain the partial dynamic structure 
factors 
\begin{eqnarray*}
S_{il}(k,z) = \sum^{2}_{j=1} G_{ij}(k,z)S_{jl}(k).
\end{eqnarray*}
Finally the partial dynamic structure factors appearing in Eq. (2) are given 
by 
\begin{eqnarray*}
S_{ij}(k,\omega) = \lim_{\epsilon \rightarrow 0} 
\frac{{\rm Re}[S_{ij}(k,z = \epsilon - i\omega)]}{\pi}.
\end{eqnarray*}
We insert the explicit expressions for $S_{ij}(k,\omega) (i,j = 1,2)$ 
in Eq. (1) to obtain the dynamic structure factors $S(k,\omega)$, which 
are in turn proportional to the differential neutron-scattering cross 
sections. \\

\section{RESULTS}

\noindent
We present results for a mixture of hard spheres of diameters 
$\sigma_{1}$ and $\sigma_{2} (\sigma_{2} > \sigma_{1})$,  
number densities $n_{1}$ and $n_{2}$ and of total packing fraction 
$\eta = \frac{\pi}{6} [n_{1}\sigma^{3}_{1} + n_{2}\sigma^{3}_{2}] = 0.46$.
The diameter ratio $\sigma_{1}/\sigma_{2}$ is taken to be 0.7. 
We vary the concentration $x_{2} = n_{2}/(n_{1} + n_{2})$ of larger 
spheres keeping $\eta$ constant. We have chosen the ratio $b$ of the 
scattering lengths of the two species to be 0.7. Finally we consider 
mixtures of spheres of equal masses. This is done mainly to focus on the 
role of structural parameters in slowing down the dynamics.\\ 

\noindent
Figures 1 and 2 show $S^{*}(k,\omega)$ [i.e., $S(k,\omega)$ scaled with 
its value at $\omega = 0$] as a function 
of the frequency $\omega$ scaled with a time $t_{E}$ given by 
$\frac{1}{t_{E}} = \frac{4\sqrt{\pi n\sigma^{2}_{12} \xi_{12}}}{\sqrt{2\beta\mu_{12}}}$, with $n$ the total number density of the fluid, 
$\sigma_{12} = \frac{\sigma_{1} + \sigma_{2}}{2}, \xi_{12}$  
the pair correlation function at contact between type 1 and type 2 spheres and 
$\mu_{12} = m_{1} m_{2}/(m_{1} + m_{2})$ the reduced mass. All plots 
of $S^{*}(k,\omega)$ have been made for $k\sigma_{12} = 2\pi$. We notice 
that in Fig. 1, the curve for $x_{2} = 0.01$ is significantly narrower than 
that for $x_{2} = 0.9$. This 
observation can be interpreted as follows. A dense liquid mixture 
consisting of a few large particles suspended in a liquid of small 
particles slows down the dynamics on molecular length scales more easily 
than a liquid with a few small particles in 
the background of large ones. This is due to the fact that large spheres 
diffuse very slowly in a matrix of small ones. In contrast, the diffusion 
coefficient of small spheres being large, they diffuse faster in a 
background of large ones. This observation agrees with recent molecular 
dynamic simulations in dense mixtures [7], where the authors find that a 
liquid mixture with a majority of large spheres is easier to crystallize 
compared to one which consists mainly of small spheres. This stems from the 
difference in the rate of diffusion of the two species as mentioned above.

%

\noindent
In Fig. 2 we notice that a compositionally disordered mixture $(x_{2} = 0.2)$
characterized by nearly equal packing fractions 
($\eta_{1} = 0.58\eta , \eta_{2} = 0.42\eta$) of the two species, 
undergoes a dramatic narrowing reflecting the  
emergence of a very slow dynamical process. In contrast, dynamical 
relaxation is relatively faster in a mixture consisting mainly of large 
spheres ($x_{2} = 0.9$). In a disordered liquid mixture, due to a difference 
in the sizes of the two species, the dynamics is slowed down considerably. 
This happens because both small as well as large spheres can be trapped 
in a cage formed by the other particles, resulting in the formation of 
peaks in $S_{ij} (k) (i,j = 1,2)$ at well separated values of the wave 
vector. In contrast, in a liquid consisting mainly of large spheres, 
caging is relatively ineffective due to the ease of movement of small 
particles in the matrix of large ones. \\

\section{CONCLUSION}

\noindent
We have obtained the dynamic structure factors of dense mixtures on 
molecular scales by confining ourselves to the modes of total density 
fluctuations and interdiffusion, which govern the slow dynamics on such 
length scales. The relaxation of density fluctuations on molecular scales 
depends more strongly on the structural parameters, such as the size 
ratio, than on the mass ratio of the two species [5,8]. This is a 
consequence of the crucial dependence of 
the dynamics on the rigid static structure of a dense liquid on molecular 
scales. Therefore, we have considered mixtures of equal masses but different 
{\it sizes}. We have confined ourselves to mixtures of not too disparate 
scattering lengths in order to have comparable contributions from the 
two species to the total $S(k,\omega)$.\\

\noindent
{\it Comparison with earlier work}: Our present analysis is more complete 
compared to previous work [4] in that 
it takes into account effects of interdiffusion, wave-vector dependence of 
transport coefficients, and incorporates a few additional static couplings 
in the extended hydrodynamic equations which had been neglected earlier. 
We have checked that our results converge to that of [4] in the appropriate 
limit. We would like to point out that the half-widths evaluated from our 
theory differ from those predicted by [4]. For instance for $x_{2} = 0.5$ 
we find that the half-width predicted by [4] is about 1.5 times that 
obtained from the present theory. This observation points to the fact 
that inclusion of the additional effects incorporated in the present 
analysis makes a significant difference to the dynamic structure factor 
of a dense mixture on molecular scales. Whether the present analysis 
improves over the earlier theory can only be settled by future 
neutron-scattering experiments.\\

\noindent
As mentioned in Sec. III, we have demonstrated that our quantitative 
evaluation of $S(k,\omega)$ confirm the qualitative observations made 
in [7] regarding the difference in the dynamics of slowing down between 
a mixture consisting of a few large particles and one containing a few 
small particles. Earlier researchers [6] had focused on neutron scattering 
in dilute to moderate density ($\eta << 0.46$) fluid mixtures. They had 
mainly concentrated on unusual sound propagation in such systems. The 
regime of density considered here is quite different. We have looked at 
high density mixtures where structural relaxation is important and the 
sound modes are overdamped. Our results are applicable to dense liquid 
mixtures below the glass transition packing fraction.\\

\section{ACKNOWLEDGMENTS}

\noindent
It is a pleasure to thank M. C. Marchetti for suggesting the problem and for 
her critical comments. I have benefited from stimulating discussions with 
A. K. Sood. This work has been supported by the National Science 
Foundation under Contract No. DMR-91-12330 and the Jawaharlal Nehru 
Centre for Advanced Scientific Research and was begun at the Department 
of Physics, Syracuse University. \\

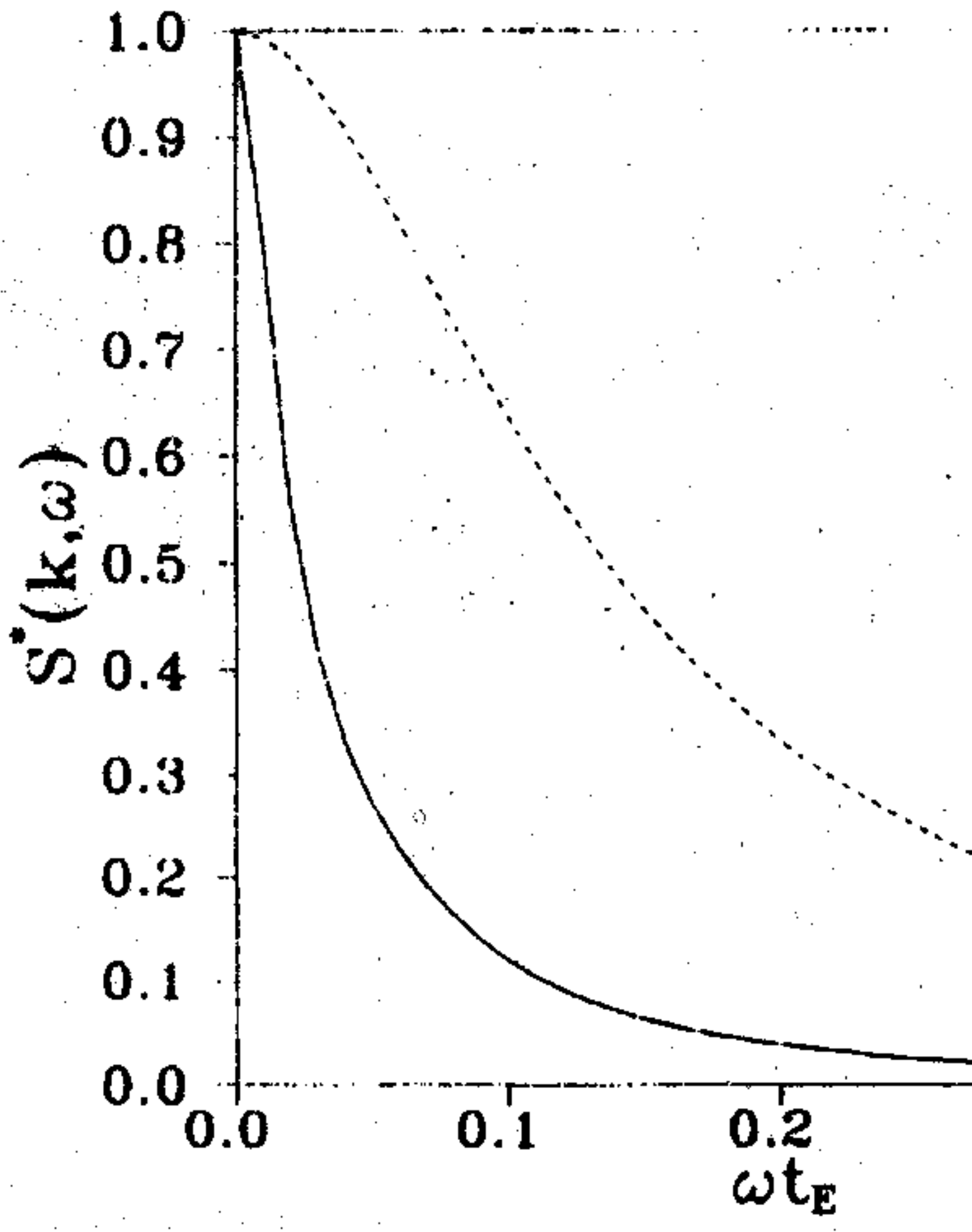
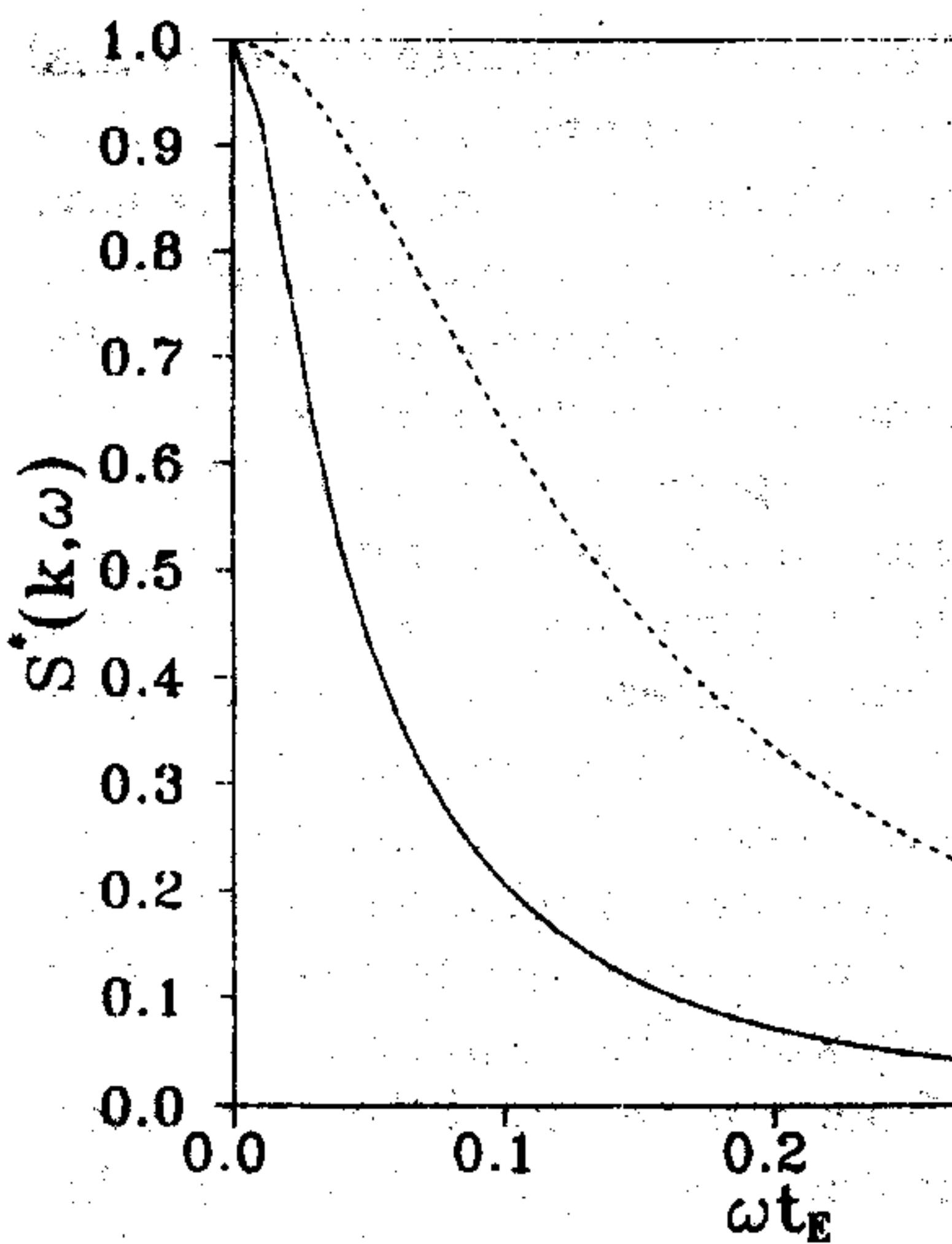
\end{document}